\documentclass{mem}
\usepackage{natbib}\usepackage{txfonts}\usepackage{balance}
\usepackage{graphicx}
\usepackage[a4paper,breaklinks]{hyperref}
\idline{0}{1}
\begin{document}
\def\teff{$T\rm_{eff }$}
\def\kms{$\mathrm {km s}^{-1}$}

\title{
Gaia, Non-Single Stars, Brown Dwarfs, and Exoplanets
}

   \subtitle{}

\author{
A. \,Sozzetti\inst{1} 
          }

  \offprints{A. Sozzetti}

\institute{
Istituto Nazionale di Astrofisica --
Osservatorio Astrofisico di Torino, Via Osservatorio 20,
I-10025 Pino Torinese, Italy. 
\email{sozzetti@oato.inaf.it}
}

\authorrunning{Sozzetti }

\titlerunning{Gaia, NSS, BDs, and Exoplanets}

\abstract{In its all-sky survey, Gaia will monitor astrometrically and photometrically millions of main-sequence stars with 
sufficient sensitivity to brown dwarf companions within a few AUs from their host stars and to transiting brown dwarfs on very short 
periods, respectively. Furthermore, thousands of detected ultra-cool dwarfs in the backyard of the Sun will have direct (absolute) distance 
estimates from Gaia, and for these Gaia astrometry will be of sufficient precision to reveal any orbiting companions with masses as low 
as that of Jupiter. Gaia observations thus bear the potential for critical contributions to many important questions in brown dwarfs 
astrophysics (how do they form in isolation and as companions to stars? Can planets form around them? What are their fundamental parameters 
such as ages, masses, and radii? What is their atmospheric physics?), and their connection to stars and planets. The full legacy potential 
of Gaia in the realm of brown dwarf science will be realized when combined with other detection and characterization programs, both from the ground and in space. 

}
\maketitle{}

\section{Introduction}

The age of micro-arcsecond ($\mu$as) astrometry has finally dawned upon us. 
ESA's global astrometry mission Gaia was successfully launched from the Kourou 
site in French Guyana on December 19, 2013. It is now concluding its commissioning 
phase after injection in its L2 Lissajous orbit, and will commence science operations 
in Summer 2014 (for details, see de Bruijne, this volume). 

\begin{figure*}[t!]
\includegraphics[width=1.0\textwidth]{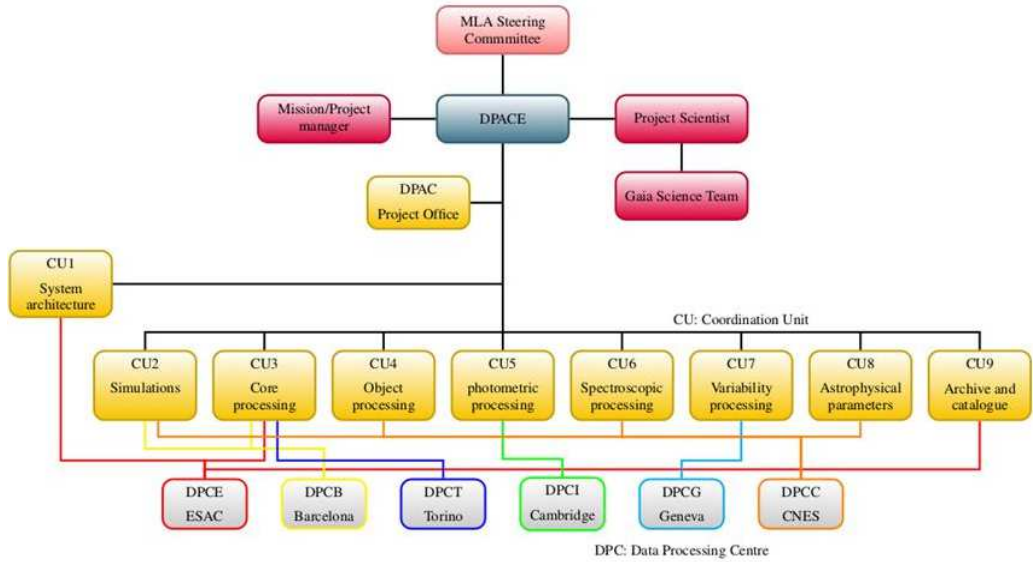}
\caption{Overall structure of the Gaia Data Processing and Analysis Consortium (DPAC) from \url{http://www.cosmos.esa.int/web/gaia/dpac/consortium}. 
This large pan-European team of expert scientists and software developers is responsible for the processing of Gaia's data with the final 
objective of producing the Gaia Catalogue. DPAC is composed of more than 450 members from over 20 countries.}
\label{dpac}
\end{figure*}

The wealth of high-quality data Gaia will collect in its magnitude-limited 
($V\leq 20$ mag) all-sky astrometric survey (complemented by onboard photometric and 
partial spectroscopic information) is bound to revolutionize our understanding of 
countless aspects of astronomy and astrophysics within our Milky Way, and beyond (e.g.,~\citealt{perryman01}). 
One key element for much improved understanding of star and planetary systems formation and evolution processes will be 
Gaia's ability to provide astrometric and spectro-photometric information for thousands of ultra-cool (brown) dwarfs in the Solar 
neighborhood, either directly detected or identified as companions to stars. 
In this paper I describe the elements of the Gaia Data Processing and Analysis Consortium (DPAC) software chain that will deal with deriving astrometric 
orbits for non-single stars within the realm of Coordination Unit 4 (Object Processing), with a focus on the strengths and challenges
of the chosen approach. I then outline some of the key science topics that might be addressed using the 
potential harvest of astrometrically detected brown-dwarf companions to normal stars and planetary-mass companions to brown dwarfs.

\section{Gaia Non-Single-Star Processing}

\begin{figure*}[t!]
\includegraphics[width=1.0\textwidth]{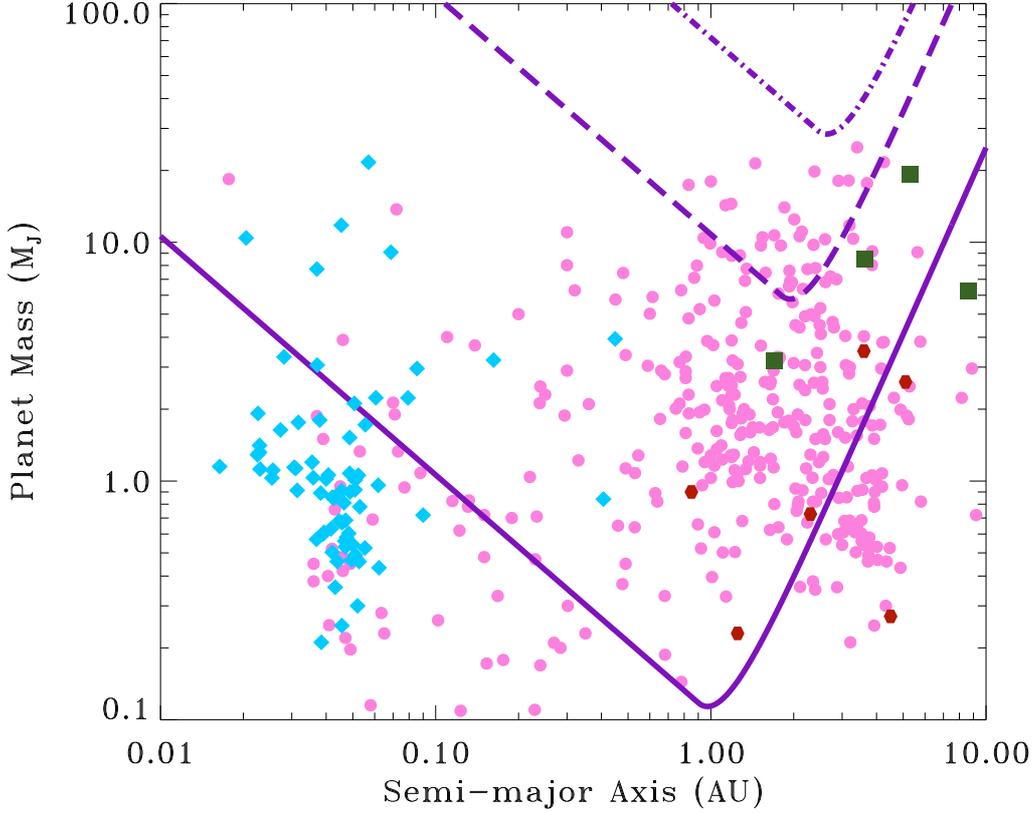}
\caption{Gaia brown dwarfs discovery space (purple curves). 
Detectability curves are defined on the basis of a 3-$\sigma$ criterion for signal detection (see~\citealt{sozzetti10} for details).
The upper dashed-dotted and center dashed curves are for Gaia astrometry with $\sigma_\mathrm{A} = 120$ $\mu$as,
assuming a 0.8-$M_\odot$ primary at 300 pc and for $\sigma_\mathrm{A} = 400$ $\mu$as, assuming a 0.2-$M_\odot$ primary at 30 pc, respectively. 
The lower solid curve is for $\sigma_\mathrm{A} = 500$ $\mu$as, assuming a 0.050-$M_\odot$ primary at 2 pc (appropriate for Luhman 16A). 
The survey duration is set to 5 yr. The pink filled circles indicate a representative samples of Doppler-detected exoplanets. Transiting systems
are shown as light-blue filled diamonds. Red hexagons are planets detected by microlensing. Planets detected with the timing 
technique are also shown as green squares. }
\label{discovery}
\end{figure*}

Fig.~\ref{dpac} provides an overview of the DPAC structure, built around specialized sub-units known as Coordination Units (CUs). 
Each CU is responsible for developing the scientific algorithms and software corresponding to a particular sub-system of the overall 
Gaia data processing system. The software developed by the CUs is run by one of the data processing centres (DPCs). 
These centers host the computing hardware and provide software engineering expertise to support the CU software development work. 
Each DPC supports at least one CU. In particular, CU4 will further process any ill-behaved objects that might not be described 
simply in terms of a five-parameters astrometric model by CU3's Astrometric Global Iterative Solution (AGIS. See e.g.~\citealt{lindegren12}), 
that might be found photometrically and/or spectroscopically variable by CU5 and CU6 and or recognized as potentially eclipsing systems by CU7. 
Such objects include Non-Single Stars (NSS), Solar System Objects (SSO), and Extended Objects (EO). 

CU4 will tackle the NSS problem by attempting to derive, based on the available spectroscopic and photometric Gaia data, spectroscopic and/or eclipsing binary solutions. 
As for astrometry, a cascade of increasingly more complex models will describe the data in terms of solutions containing derivatives of the stellar 
proper motion, accounting for variability-induced motion, all the way to fully Keplerian astrometric orbital solutions, including where appropriate 
(particularly for the case of extrasolar planetary systems) multiple companions. It is worth underlining that for the most complex models, involving a large number of adjustable 
parameters many of which non-linear, the assessment of their reliability and robustness (including meaningful error estimates on the fitted quantities) 
will be a non-trivial task, particularly in the limit of astrometric signals comparable in size to Gaia's single-measurement uncertainties and/or 
limited redundancy in the number of observations with respect to the model parameters (see, e.g.,~\citealt{sozzetti12}).

\section{Gaia Sensitivity Estimates} 

In the recent past,~\citet{casertano08},~\citet{sozzetti14} and~\citet{dzigan12} have quantified the sensitivity of 
Gaia astrometry and photometry to giant planetary companions at intermediate and very short separations around F-G-K-M-type dwarfs. 
They found that Gaia should provide a sample of maybe $10^4$ new astrometrically detected giant planets and a few thousands transiting 
hot Jupiters. The results of the above mentioned works can be used to extrapolate the Gaia yield of brown dwarf (BD) companions to 
main-sequence stars with any orbit orientation in space. For BDs bright enough to be directly detected by Gaia, their masses being in the sub-stellar 
regime ($M\approx10-80$ $M_J$), the sensitivity of Gaia measurements might also enable detection of very low-mass companions around this sample. 

\subsection{Gaia Photometry: Eclipsing BDs}

Brown dwarfs have radii that vary by only 10-15\% over the range of possible masses, and they all roughly have the same radius as Jupiter. This is a result 
of the dominant equation of state for their interiors (classical ionic Coulomb pressure + partial electron degeneracy pressure. See e.g., \citealt{basri06}). 
The expected Gaia per-transit photometric standard errors are on the order of 0.001 mag for stars with magnitudes $G\leq13$ mag, and still in the 0.003-0.005 mag range 
at $G = 16$ mag~\citep{jordi10}. Detection of transiting Jupiter-sized objects (such as BDs) around solar-type (or later-type) stars is thus relatively easy, as the corresponding 
transit depths are on the order of 0.01 mag or more. The detection limits for close-in (a few days of period) Jupiter-sized 
companions identified by~\citet{dzigan12} thus apply directly to the potential sample of transiting BDs found with Gaia photometry. 
The most serious limitation is naturally the low-cadence of Gaia observations, however mitigated at least 
in part by the sheer numbers (millions) of stars potentially available for transit detection. 
 
\subsection{Gaia Astrometry: BDs Companions to Stars}

The astrometric detection of giant planetary companions ($M\approx 1$ $M_J$) around main-sequence stars takes full advantage of Gaia's limiting positional precision of 
$\sigma_\mathrm{A}\simeq15-20$ $\mu$as achieved uniformly for bright objects with $G\leq13$ mag (e.g.,~\citealt{sozzetti14}). The masses of BDs can be almost a factor of 10 
larger than the most massive objects defined as planets ($M\approx 13$ $M_J$) based on the simple mass-threshold (e.g.,~\citealt{oppenheimer00}) adopted in IAU conventions. 
Their detection in Gaia astrometry can thus be achieved allowing for comparatively larger values of $\sigma_\mathrm{A}$, depending on the actual combination of target mass, distance, and apparent 
brightness at G-band. The regions above the upper and center curves of Fig.~\ref{discovery} show the discovery space in the mass-orbital separation diagram for two representative 
systems. One thus infers that BD detection with Gaia can be achieved around a much larger stellar sample ($10^6-10^7$ objects) than available for giant planet discovery, depending 
on spectral type. 

\subsection{Gaia Astrometry: Planetary Companions to BDs}

At the presently envisioned faint-end limit ($G=20$ mag) of the Gaia survey, a few thousands isolated BDs will be directly observed by Gaia (see de Bruijne this volume, 
and references therein). The typical per-measurement precision in Gaia astrometry for these very faint objects will be on the order of 500-700 $\mu$as. However, the combination of 
very low masses of BDs and their being found in the close vicinity of the Sun ($d< 20-30$ pc) due to their intrinsic low luminosity will allow Gaia to detect orbiting very low-mass 
secondaries around this sample, if present in the orbital separation range to which Gaia is most sensitive. The lower curve of Fig.~\ref{discovery} shows the Gaia sensitivity limits 
in mass and separation for companions orbiting Luhman 16~\citep{luhman13}, a well-studied BD binary (the closest known BD and the third closest system to the Sun). 
One of the components appears to host an intermediate-separation low-mass (possibly planetary) companion~\citep{boffin14}. As one can see, if present, the companion would be 
clearly detectable in Gaia astrometry, that would also easily identify around which of the two components the object is actually orbiting. 

\section{Discussion}

The detection with Gaia of brown dwarfs in the three categories mentioned in the previous Section will pave the way to important advances in our comprehension of 
many issues in connection with the formation and evolution processes of stars and sub-stellar objects (BDs and planets). We highlight here some of the most relevant ones. 

\subsection{The BD - Giant Planet Connection}

The physical properties of BDs and giant planets are expected to bear important elements of similarity, in terms of mass, temperature, gravity, but also substantial differences, in terms of 
atmospheric chemistry, mechanisms of cloud formation, and general interpretation of spectral features (e.g.,~\citealt{madhusudhan14}). Close-in, strongly irradiated 
BDs and gas giant planets found in transit by Gaia could bring statistically significant samples to improve the overall understanding of these two partially overlapping 
classes of astrophysical objects. In particular, data from the CoRoT and Kepler mission allow to infer that the occurrence rate of close-in ($P\lesssim5$ days ) BDs 
is at least one order of magnitude lower than that of hot Jupiters (e.g.,~\citealt{moutou13}, and references therein). A simple extrapolation from the expected numbers of transiting hot-Jupiters uncovered 
by Gaia (\citealt{dzigan12}; see also Dzigan, this volume) around stars with $G\leq16$ mag would indicate that on the order of a few hundred transiting BD candidates 
might be identified in Gaia photometry. While follow-up efforts for mass confirmation with precision radial velocities (RVs) will still suffer some difficulties due to the typical faintness of 
the primaries (the bulk of detections expected to be at $G>14$ mag), these will definitely not be severe given the range of BD masses and thus expected RV 
signals in the ballpark of at least a few km/s (see also Bouchy this volume). 

\subsection{Characterizing the BD Desert}

The differences in giant planet and BD occurrence rates extend to a wide range of orbital separations. In particular, intermediate-separation ($a<3-4$ AU) 
BD companions to solar-type stars appear rare ($<1\%$) when compared to the fraction of giant planet hosting stars at similar separations ($\sim7\%$). 
At wider separations ($\approx20-1000$ AU) this number rises to $2-3\%$ (for a review, see~\citealt{ma14}, and references therein). 
The paucity of close-in BD companions to solar-type dwarfs is usually referred to as the `brown dwarf desert'. Furthermore, the statistical properties of BDs 
at intermediate separations from RV surveys allow to infer the existence of a particularly `dry' region at short periods and medium masses~\citep{ma14}, 
with low- and high-mass BDs exhibiting significantly different eccentricity distributions. 

All the above results bear relevant insights to improve our understanding of the different formation mechanisms for BDs (and their relative role), 
in connection to those of stars and planets. Nevertheless, the findings still suffer from small-number statistics. For example, $1-\sigma$ errors on 
BD occurrence rates from Doppler surveys have present-day best-case uncertainties of 30\% (some 60 objects known to-date). Based on the Gaia sensitivity thresholds for 
astrometric detection of BD companions to relatively bright ($G\lesssim16$ mag) F-G-K-M stars and the BD fraction inferred from RV data (a $10^4$-fold increase in 
the target sample) one then expects 1000s of detections in Gaia astrometry. Gaia thus has the potential to completely characterize the BD desert, with 
fine-structure studies of its dependence on stellar properties such as mass and metallicity. The wealth of Gaia data on BD companions to stars will thus allow 
to probe effectively possible differences in BD formation mechanisms at intermediate separations (see also Parker, this volume). 

\subsection{Planet Formation Around BDs}

Observations of accretion disks around young BDs (e.g.,~\citealt{ricci12}) have led to the speculation that they may form planetary systems similar to normal stars. Microlensing surveys have provided 
the first evidence of the existence of planetary-mass objects around BDs. Most of the detections have occurred for planetary companions to BDs at relatively large separations (\citealt{chauvin04};
~\citealt{todorov10}), thus leaving room for formation mechanisms analogous to stellar binaries for such systems. There is however one recent notable exception~\citep{han13} that 
provides initial observational support to the hypothesis that planet formation processes may not stop around sub-stellar mass primaries. 

Clearly, additional theoretical work is needed to explore the viability of planet formation in BD protoplanetary disks, either by the gravitational instability or 
core accretion mechanism. For this, it is essential to find more binaries with BD hosts in wide ranges of mass ratios and separations. In this respect, we have seen 
how Gaia will be capable of detecting a few thousands BDs of all ages (and up to a factor of 4 more if a limiting $G=21$ mag is achieved, see de Bruijne this volume), 
with sufficient astrometric sensitivity to giant planets orbiting within $2-3$ AUs. Gaia astrometric measurements of directly detected BDs thus bear the potential to provide 
a fundamental observational test of planet formation around ultra-cool dwarfs.

\section{Conclusions}

Gaia astrometry and photometry are set to provide very `cool' results in the realm of ultra-cool (brown) dwarfs in the Solar neighborhood. The wealth of Gaia data 
will 1) allow to pin down their occurrence rates as companions to normal stars at short and intermediate separations with outstanding statistical resolution (in bins of 
both primary and secondary masses), 
2) enable measurements of their radii for those found transiting the disk of their hosts, 3) deliver exquisitely precise direct distance estimates and detailed spectral characterization 
for those directly detected (see de Bruijne this volume, Sarro this volume), and 4) provide the first-ever statistically robust estimates of giant planet frequencies around them. 

The power of the Gaia legacy in brown dwarfs astrophysics is even better understood when seen as an element of synergy with other ground-based and space-borne programs 
to address many key questions related to the formation of brown dwarfs, their fundamental parameters (mass, radius, age), their atmospheric physics and chemistry, the likelihood 
that they might host planetary systems, and those unforeseen 
aspects that still need to be unveiled (see for example contributions to this proceedings volume by Bochanski, Bouchy, Dzigan, Faherty, Joergens, Kirkpatrick, and Sahlmann).

\begin{acknowledgements}

I thank the SOC and LOC of the workshop for the realization of a scientifically vibrant, smoothly run event 
(with an added value of social entertainment!). This work has been funded in part by ASI under contract to INAF 
I/058/10/0 (Gaia Mission – The Italian Participation to DPAC). This research has made use of NASA's Astrophysics Data System.

\end{acknowledgements}

\bibliographystyle{aa}

\end{document}